\documentclass[twocolumn,showpacs,preprintnumbers,amsmath,amssymb]{revtex4}

\usepackage{graphicx}

\begin{document}


\title{Dynamic positive column in long-gap barrier discharges.}



\author{A. Shvydky}
\email[{Electronic mail: }] {ashvidk@physics.utoledo.edu}
\author{V. N. Khudik$^\sharp$, V. P. Nagorny$^\sharp$}
\author{C. E. Theodosiou}

\affiliation{{Department of Physics and Astronomy, University of
Toledo, Toledo, OH 43606 USA}}

\affiliation{{$^\sharp$Plasma Dynamics, Corp., 417 E. 8 Mile Rd,
Hazel Park, MI 48030 USA}}



\date{\today}

\begin{abstract}
A simple analytical model of the barrier discharge in a long gap
between opposing plane electrodes is developed. It is shown that the
plasma density becomes uniform over large part of the gap in the
course of the discharge development, so that one can speak of a
formation of a dynamic positive column. The column completely
controls the dynamics of the barrier discharge and determines such
characteristics as the discharge current, discharge duration, light
output, etc. Using the proposed model, all discharge parameters can
be easily evaluated.
\end{abstract}



\maketitle


\section{\label{sec:intro}INTRODUCTION}
Dielectric barrier discharges are typically generated in gas gaps
between two electrodes covered with dielectric layers. Sinusoidal or
square-wave voltage with frequencies ranging from few kHz to
hundreds of kHz are used to generate the discharges. The product of
the gas pressure and the distance between the dielectric surfaces $p
D$ can vary quite significantly. In the present paper we consider
barrier micro-discharges such as used in Plasma Display Panels
\cite{Boeuf95} and excimer lamps \cite{Eliasson91} where $p D
\approx 5-50$.

The dynamics of a barrier discharge during one pulse of the applied
square-wave voltage was considered in detail analytically in our
previous work \cite{Khudik2003}, where it was shown that at high
overvoltage the discharge develops into an ionization wave, whose
velocity is determined primarily by the charge production rate in
the cathode fall region. This wave moves from the anode toward the
cathode, resulting in contraction of the cathode fall region and
increase of the electric field within this region. Upon reaching the
cathode, the ionization wave can either quickly disappear (when the
capacitance of the dielectric layer is small) or transform into a
quasi-stationary DC cathode fall (when the capacitance is large).
The main assumption of that work was that the resistance of the
plasma trail, created by the ionization wave, can be neglected.

\begin{figure}
    \includegraphics[width=2.5in]{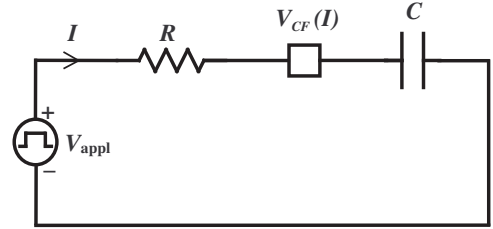}
    \caption{\label{fig:ZenerCircuit} Equivalent circuit of a long-gap
barrier discharge.}
\end{figure}

In the present paper we consider the opposite case, when the barrier
discharge dynamics is strongly influenced by the plasma trail which
electrically connects the anode with the cathode fall region. As
will be shown below, in the case of a long gap the plasma trail
becomes uniform over the large part of the gap, virtually forming a
dynamic positive column. The uniformity of the column enables us to
consider it simply as a variable resistor through which the cathode
fall (CF) charges the dielectric layer capacitor (see
Fig.~\ref{fig:ZenerCircuit}). When the capacitance is large, one can
introduce further simplifications into the model. In this case the
CF is, in essence, quasi-stationary and its V-I characteristics can
be approximated by those of the DC cathode fall \cite{RaizerBook}.

In our work we neglect the volume and near-wall recombination of
charged particles, which if included into consideration would only
increase the resistance of the positive column and enhance its
role in the barrier discharge dynamics.

Although dynamic characteristics of barrier discharges have been
already extensively studied via computer simulations (see
Ref.~\cite{Boeuf2003review} and references therein), it remained
unclear, for example, what processes control the duration and
amplitude of the current pulse. The proposed model answers this
question and can be used to estimate various parameters of long-gap
barrier discharges between opposing plane electrodes. This type of
discharges (as well as
\cite{Weber2001patent,JerrySID2000,Kawai2004}) may prove to be a
viable alternative to near-surface discharges
\cite{Boeuf99,Kushner99I,Khudik2005SID,Khudik2005Images} favored
nowadays in Plasma Display Panel industry.

\section{\label{sec:general}Qualitative consideration and basic equations}

We will consider discharges in a gas gap between opposing plane
electrodes covered by dielectric layers of thickness $d$ and
dielectric constant $\epsilon$. We assume that the gap length $L_g$
is much larger than the length of the normal cathode fall,
\begin{equation}
            L_g \gg L_{\textrm{norm}}.
            \label{eq:LgLcf}
\end{equation}
It is also assumed that the dielectric layer capacitance is large,
i.e. the effective thickness of dielectric layers  is small,
\begin{equation}
            2 d/\epsilon \lesssim L_{\textrm{norm}}.
            \label{eq:LdielLcf}
\end{equation}
 For example, in the case of 10\%-Xe/90\%-Ne mixture at pressure
500 Torr and secondary emission coefficients $\gamma_{Ne}=0.5$ and
$\gamma_{Xe}=0.005$, the normal cathode fall length
$L_{\textrm{norm}}\sim 6.5 \mu m$, the gap length under
consideration is from several hundred microns to one millimeter, and
the effective thickness of dielectric layers is less than or about
several microns.

Although discharges in gaps with such a long distance between the
anode and cathode can be initiated in several different ways (by,
for example, using a set of auxiliary electrodes), we assume that
the Townsend breakdown takes place .

Under these assumptions (long gap, large capacitance, Townsend
mechanism of the breakdown), the general picture of the discharge
dynamics can be described as follows:

1. Upon application of a sufficiently high voltage across the gap
($U_{\textrm{appl}}>U_{\textrm{br}}$, where $U_{\textrm{br}}$ is the
breakdown voltage), the Townsend criterion is satisfied and the
positive charge starts to build up in the gap. While the amount of
total charge is small (and thus the electric field is undisturbed),
it grows exponentially with time.

2. At some moment it reaches a critical value ($\sim\epsilon_0
V_{\textrm{appl}}/L_g$) and causes considerable distortion of the
electric field. The field vanishes at the anode and from this moment
on there coexist two different regions in the gap: the region filled
with plasma (plasma trail) where the electric field is relatively
small, and the region of the gap adjacent to the cathode where the
electric field is strong and the electron density is negligible.
With time, the plasma trail expands toward the cathode while the CF
region contracts. At this stage, the process of the plasma trail
uniformization begins.

\begin{figure}
    \includegraphics[width=3.0in]{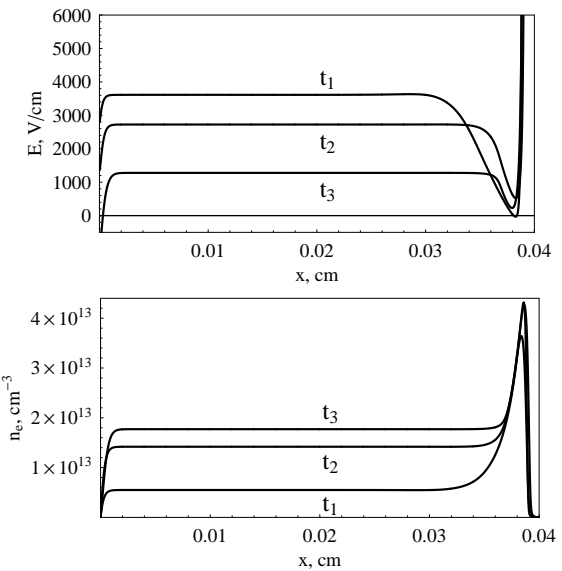}
    \caption{\label{fig:StageII} Electric field $E$ and electron density $n_e$ in the
        gap at time moments $t_1<t_2<t_3$ when the discharge current is
        strong: $I(t_1)=0.5 I_\textrm{max}$, $I(t_2)=I_\textrm{max}$, and $I(t_3)=0.5 I_\textrm{max}$,
        where $I_\textrm{max}$ is the maximum discharge current.}
\end{figure}

3. When the length of the CF region becomes comparable to
$L_{\textrm{norm}}$, it transforms into a quasi-stationary DC
cathode fall and the discharge current sharply increases and, after
reaching its maximum, decreases. The plasma density in the already
uniformized plasma trail (positive column) grows while the electric
field there falls (see Fig.~\ref{fig:StageII}).
These processes are accompanied by the intensive deposition of power
from the external source into the discharge. At the end of this
stage, the main portion of the external voltage drops across the
dielectric layer capacitor, which leads to quenching of the
discharge.

4. In the afterglow, the number of ions and electrons in the gap
gradually decreases through the dissociative recombination of
electrons and molecular ions. Charged particles are also pulled out
of the plasma toward the dielectric surfaces by the residual (and
ambipolar) electric fields. Note that although the discharge current
is small during the discharge decay, the voltage transferred to the
capacitor is about the normal CF voltage $V_{\textrm{norm}}$ (when
$2 d/\epsilon \ll L_{\textrm{norm}})$.

The key feature of long-gap barrier discharges is that the plasma
trail is uniform during the stage three, when the discharge current
is strong. The intrinsic mechanism responsible for the
uniformization can be qualitatively explained as follows: Due to the
plasma quasi-neutrality, which quickly establishes in the Maxwellian
time (see Sec.~\ref{sec:discussion}), the particle current is
constant throughout the plasma trail. Therefore the electric field
is higher in regions where plasma density is lower. The higher
electric field results in higher ionization rate, which in turn
leads to faster growth of the plasma in these regions and,
eventually, to a leveling off of the plasma density.

To find the plasma density in the positive column, one can use the
continuity equation for electrons
\begin{equation}
            \frac{\partial n}{\partial t}-\frac{\partial n \mu_e E}{\partial
            x}=r(E)\,n,
            \label{eq:dnedt}
\end{equation}
where $n$ is the electron (plasma) density, $\mu_e$ is the electron
mobility, $E$ is the electric field in the column, and $r(E)$ is the
ionization rate. For simplicity, we omitted the diffusion part of
the electron flux in Eq.~(\ref{eq:dnedt}). Neglecting the motion of
ions (ion mobility is several hundred times lower than the electron
mobility) and using the constancy of the particle current, one can
immediately arrive at the following relationship between the plasma
density and the electric field in the positive column,
\begin{equation}
            e n \mu_e E=I(t)/S,
            \label{eq:mueEnj}
\end{equation}
where $e$ is the elementary charge, $I(t)$ is the discharge current,
and $S$ is the cross-sectional area of the discharge. Since the
electron mobility is almost constant at small electric fields, from
Eq.~(\ref{eq:mueEnj}) it follows that
\begin{equation}
            \delta n E_{pc} + n_{pc} \delta E =0,
            \label{eq:dnEpc}
\end{equation}
where $\delta n$ and $\delta E$ are perturbations and $n_{pc}$ and
$E_{pc}$ are the average values of the plasma density and electric
field in the plasma column. Using
Eqs.~(\ref{eq:dnedt})-(\ref{eq:dnEpc}), one can readily obtain
\begin{eqnarray}
            \frac{\partial n_{pc}}{\partial t}=r(E_{pc})\,n_{pc},
            \label{eq:dndtpc} \\
            \frac{\partial \delta n}{\partial t} = \Bigg[ r(E_{pc}) - E_{pc} \frac{d r(E_{pc})}{d E_{pc}} \Bigg] \; \delta
            n.
            \label{eq:ddeltandtpc}
\end{eqnarray}
While the obvious equation (\ref{eq:dndtpc}) describes growth of the
plasma density in the positive column, Eq.~\ref{eq:ddeltandtpc}
shows fast decay of the density perturbations (since under positive
column conditions $E_{pc} r'(E_{pc}) > r(E_{pc})$). Let us emphasize
that, as is seen from our derivation, the mechanism of the positive
column uniformization has an entirely dynamic nature and manifests
itself regardless of the diffusion processes. It is interesting to
note that the uniformization takes place even in a more general case
\cite{shvydky2004} when the discharge can not be treated as
quasi-one-dimensional.

In this paper we give a simple analytic description of only the
third stage of the discharge development, when discharge current is
strong. During this stage, the positive column occupies almost the
entire gap, and as seen from Eq.~(\ref{eq:dndtpc}) its resistance $R
\approx L_{g}/(e n_{pc} \mu_{e} S)$ is governed by
\begin{equation}
            \frac{d R}{d t}=-r(E_{pc})\, R,
            \label{eq:dRdt}
\end{equation}
where
\begin{equation}
            E_{pc}=\frac{I R}{L_g}.
            \label{eq:Epc}
\end{equation}
The quasi-stationarity of the cathode fall allows us to use the V-I
characteristic of the corresponding DC cathode fall,
\begin{equation}
            V_{CF}=V_{CF}(I).
            \label{eq:Ucfofj}
\end{equation}
The applied voltage is distributed between the elements of the
circuit in Fig.~\ref{fig:ZenerCircuit} according to the second
Kirchoff law,
\begin{equation}
            I R + V_{CF}(I) + \frac{q}{C}= V_{\textrm{appl}},
            \label{eq:KirchoffLaw}
\end{equation}
where $C=\epsilon_0 \epsilon S/2 d$ is the capacitance of the
dielectric layer capacitor, and the capacitor charge $q$ changes
with time as
\begin{equation}
            \frac{d q}{d t}=I.
            \label{eq:dqdtI}
\end{equation}

The equations (\ref{eq:dRdt})-(\ref{eq:dqdtI}) form the foundation
of our model of barrier discharges with long positive columns and a
quasi-stationary cathode fall. They must be complemented with the
following initial conditions
\begin{eqnarray}
            R(t \rightarrow -\infty) =\infty,
            \label{eq:npinit}\\
            q(t \rightarrow -\infty) =0.
            \label{eq:qinit}
\end{eqnarray}

\begin{figure}[h!]
    \centering
    \includegraphics[width=3in]{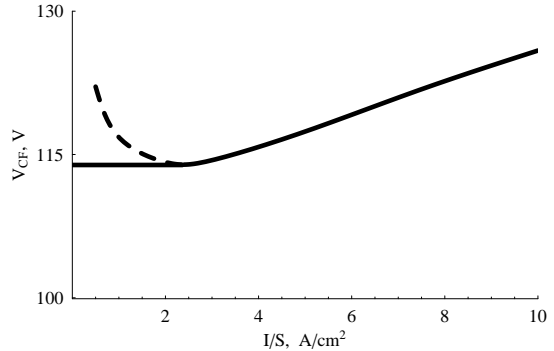}
    \caption{\label{fig:VIofCFnew}V-I characteristics of the DC cathode
    fall obtained in the fluid approximation for the gas
    mixture and secondary emission coefficients mentioned at the
    beginning of Sec.~\ref{sec:general}. Dotted curve corresponds to the unstable subnormal cathode fall.}
\end{figure}

Note, that at not very high current densities,
the DC CF voltage changes insignificantly (see
Fig.~\ref{fig:VIofCFnew}). The case when the $V_{CF}$ is taken
constant and equal to $V_{\textrm{norm}}$ is considered in the next
section.

\section{\label{sec:constCF}Constant CF voltage approximation}

When $V_{CF}=V_{\textrm{norm}}=\textrm{const}$, it is convenient to
use $R$ and $E_{pc}$ as independent variables. Replacing the $I R$
with $E_{pc} L_{g}$ and after that taking the time derivative of
Eq.~(\ref{eq:KirchoffLaw}), one can obtain the equation for the
electric field in the positive column,
\begin{equation}
            \frac{d E}{d t}=-\frac{E}{R\, C},
            \label{eq:dEdt}
\end{equation}
where the subscript ``pc'' is omitted for brevity.
Eqs.~(\ref{eq:dRdt}) and (\ref{eq:dEdt}) must be solved with the
following initial conditions
\begin{eqnarray}
            R(t \rightarrow -\infty) = \infty,
            \label{eq:InitCondR}\\
            E(t \rightarrow -\infty) =  E_0 \equiv
            \frac{V_{\textrm{appl}}-V_{CF}}{L_g}.
            \label{eq:InitCondE}
\end{eqnarray}
Condition (\ref{eq:InitCondR}) indicates that at the beginning,
there is no plasma in the discharge gap, and condition
(\ref{eq:InitCondE}) directly follows from condition
(\ref{eq:qinit}) and the Kirchoff's law (\ref{eq:KirchoffLaw}).

We would like to point out that, based on what was said in the
previous section, Eqs.~(\ref{eq:dRdt}) and (\ref{eq:dEdt}) do not
describe the initial stages of the discharge development, when the
positive column is being formed, neither do they describe the final
stage, when the discharge quenches and the cathode fall dies.

\begin{figure}
    \includegraphics[width=2.5in]{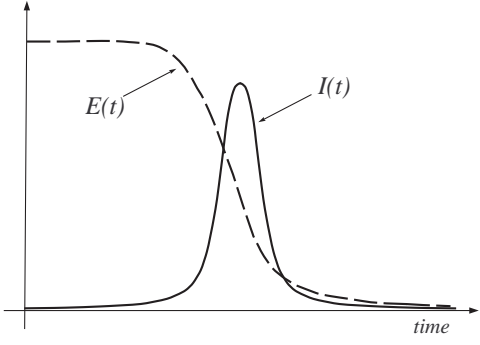}
    \caption{\label{fig:CurrentPulse} Typical dependence of $I$ and $E$ on time.}
\end{figure}

\begin{figure}[h!]
    \includegraphics[width=2.5in]{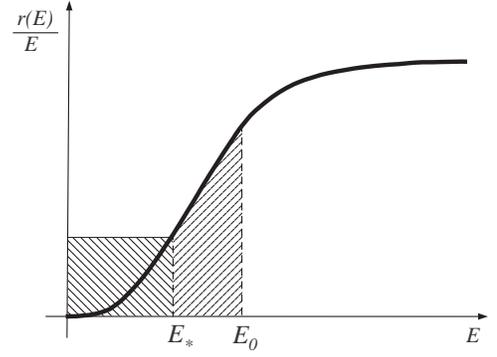}
    \caption{\label{fig:Ejmaxsolution} Graphical solution of equation~(\ref{eq:Ejmax}).
                Area of the dashed rectangle on the left is equal to
                $r(E_*)$, while the area of the dashed region under the $r(E)/E$
                curve on the right is equal to $\bar{r}(E_*,E_0)$.
                These areas are equal when $E_*$ is the electric field at the current maximum.}
\end{figure}

Now let us turn our attention to the analysis of the system of
Eqs.~(\ref{eq:dRdt}) and (\ref{eq:dEdt}). Dividing left and right
parts of equation (\ref{eq:dRdt}) by the corresponding parts of
equation (\ref{eq:dEdt}) and multiplying both parts by $C$ we obtain
\begin{equation}
            \frac{d (R C)}{d E}=\frac{r(E)}{E}\ (R C)^2.
            \label{eq:dRCdE}
\end{equation}
Integrating this equation with initial conditions
(\ref{eq:InitCondR}) and (\ref{eq:InitCondE}) gives
\begin{equation}
            \frac{1}{R C}=\bar{r}(E,E_0),
            \label{eq:oneoverRC}
\end{equation}
where
\begin{equation}
            \bar{r}(E,E_0) \equiv \int_E^{E0} \frac{r(E)}{E}\, dE.
            \label{eq:Rdef}
\end{equation}
Substitution of (\ref{eq:oneoverRC}) into (\ref{eq:dEdt}) leads to a
differential equation for the electric field,
\begin{equation}
            \frac{d E}{d t}=-E\; \bar{r}(E,E_0),
            \label{eq:dEdtfinal}
\end{equation}
which should be solved with the initial condition
(\ref{eq:InitCondE}). From this equation, one can see that the time
evolution of the electric field in the neutral column depends only
on its initial value and the ionization rate, and does not depend
explicitly on the electron mobility. Because the right part of
equation (\ref{eq:dEdtfinal}) is always negative, the electric field
monotonically decreases with time from $E_0$ to zero. The discharge
current in our model is uniquely determined by the electric field,
\begin{equation}
            I\equiv \frac{E L_g}{R} = L_g C E\;\bar{r}(E,E_0).
            \label{eq:j}
\end{equation}
It goes from zero (at $E=E_0$), trough the maximum (at $E=E_*$), to
zero again (at $E=0$), see Fig.~\ref{fig:CurrentPulse}.
The electric field $E_*$ at the current maximum can be determined
from the extremum condition $d I(E)/d E=0$, which, with the use of
Eq.~(\ref{eq:j}), can be written as
\begin{equation}
            r(E_*)=\bar{r}(E_*,E_0).
            \label{eq:Ejmax}
\end{equation}
The typical shape of the function $r(E)/E$ (which is proportional to
the the ionization coefficient $\alpha(E)$) as well as the graphical
solution of equation (\ref{eq:Ejmax}) is given in
Fig.~\ref{fig:Ejmaxsolution}.
From this figure, one can see, in particular, that always
$E_0/2<E_*<E_0$. If $r(E)/E$ grows faster than the linear function,
then $E_0/\sqrt{3}<E_*<E_0$.

Using Eqs.~(\ref{eq:j}) and (\ref{eq:Ejmax}), the expression for
maximum current can be written as
\begin{equation}
            I_{\textrm{max}}= C L_g E_*\, r(E_*).
            \label{eq:jmax}
\end{equation}
This equation shows that in the present model the maximum current is
proportional to the dielectric layer capacitance. Dependence on the
gap length $L_g$ is more complicated, since $E_*$ also depends on
$L_g$ (through $E_0$).

The time duration $\tau_{\textrm{cur}}$ of the current pulse can be
defined as the ratio of transferred charge $Q_{\textrm{tr}}=C\,
(V_{\textrm{appl}}-V_{CF})= C L_g E_0$ to the maximum current,
\begin{equation}
            \tau_{\textrm{cur}} \equiv \frac{Q_{\textrm{tr}}}{I_{\textrm{max}}}=\frac{E_0}{E_*\,
            r(E_*)}.
            \label{eq:tauj}
\end{equation}
It is useful to note that when $r(E)/E$ grows faster than the linear
function, then
\begin{equation}
            \tau_{\textrm{cur}} > \frac{3 \sqrt{3}}{r(E_0)} \approx
            \frac{5.2}{r(E_0)}.
            \label{eq:taulin}
\end{equation}
For a steeper dependence of the ionization coefficient on the
electric field, the numerical coefficient in Eq.~(\ref{eq:taulin})
is even greater.

From (\ref{eq:oneoverRC}) immediately follow the expressions for the
final resistance of the positive column,
\begin{equation}
            R(t=\infty)=\frac{1}{C \, \bar{r}(0,E_0)},
            \label{eq:Rfin}
\end{equation}
and for the final plasma density in the column,
\begin{equation}
            n(t=\infty)=\frac{C\, L_g }{e \mu_e S}\, \bar{r}(0,E_0).
            \label{eq:nncfin}
\end{equation}
Using the expression for the ionization coefficient $\alpha=r/\mu_e
E$, the ratio of the total number of electrons (ions) created in the
positive column $N=n L_g S$ to the number of transferred electrons
(ions) $Q_{\textrm{tr}}/e$ can be written in the form
\begin{eqnarray}
            \frac{e N}{Q_{\textrm{tr}}}= \langle \alpha \rangle L_g,
            \label{eq:Qnc}
\end{eqnarray}
where
\begin{eqnarray}
            \langle \alpha \rangle= \frac{1}{E_0} \int_0^{E0} \alpha(E)\,
            dE. \nonumber
\end{eqnarray}
Note that $\langle \alpha \rangle < 0.5 \alpha(E_0)$ if $\alpha(E)$
grows faster than the linear function. If, in addition to that,
$V_{\textrm{appl}} \leq V_{\textrm{br}}+V_{\textrm{norm}}$, then
there is a practically useful estimate for the number of charged
particles created in the positive column,
\begin{eqnarray}
            e N \lesssim \frac{1}{2} \alpha(E_0) L_g Q_{\textrm{tr}} <
            \frac{1}{2} \ln\big(1+\frac{1}{\gamma}\big) C V_{\textrm{appl}}.
            \label{eq:Nestimate}
\end{eqnarray}
This expression gives reasonable estimates far beyond the
applicability of our model. Note that since the electric field is
small in the positive column, in the case of noble gas mixtures,
mainly the ions of species with the lower ionization potential are
created there.

Similar formulae can be obtained for the number of excited atoms
(molecules) created in the positive column during the discharge,
\begin{eqnarray}
            \frac{e N_{\textrm{exc}}}{Q_{\textrm{tr}}}= \langle \alpha_{\textrm{exc}} \rangle L_g,
            \label{eq:Qnc}\\
            \langle \alpha_{\textrm{exc}} \rangle= \frac{1}{E_0} \int_0^{E0} \frac{r_{\textrm{exc}}(E)}{\mu_e E}\,
            dE, \nonumber
\end{eqnarray}
where $r_{\textrm{exc}}$ is the excitation rate. Again, in the case
of noble gas mixtures, mostly the species with lower excitation
energies get excited.

Using Eqs.~(\ref{eq:dEdtfinal}) and (\ref{eq:j}), the energy
deposited into the positive column can be estimated as
\begin{equation}
            W_{pc}=\frac{1}{2} C (V_{\textrm{appl}}-V_{CF})^2,
            \label{eq:Wpc}
\end{equation}
while the total energy which goes from the external source into the
discharge is
\begin{equation}
            W_{\textrm{tot}}=\frac{1}{2} C V_{\textrm{appl}}^2.
            \label{eq:Wtot}
\end{equation}
Note, that the same amount of energy, $C V_{\textrm{appl}}^2/2$, is
stored in the capacitor and is recovered during the next discharge
pulse.

The difference between Eqs.~(\ref{eq:Wtot}) and (\ref{eq:Wpc}) gives
the amount of energy that goes (in the considered approximation)
into the cathode fall,
\begin{equation}
            W_{CF}=\frac{1}{2} C \big[V_{\textrm{appl}}^2-(V_{\textrm{appl}}-V_{CF})^2\big].
            \label{eq:WCF}
\end{equation}

\section{\label{sec:discussion}Discussion}

 To verify the validity of the approximation $V_{CF}=V_{\textrm{norm}}$,
one should substitute $I_{\textrm{max}}$ from Eq.~(\ref{eq:jmax})
into the \mbox{V-I} characteristics of the DC cathode fall
(\ref{eq:Ucfofj}) and make certain that
\begin{equation}
            V_{CF}(I_{\textrm{max}}) \approx V_{\textrm{norm}}.
\end{equation}
If the obtained CF voltage is noticeably higher than the normal CF
voltage, one should analyze the full system of
Eqs.~(\ref{eq:dRdt})-(\ref{eq:dqdtI}), i.e. use the more accurate
$V_{CF}=V_{CF}(I)$ approximation instead of the
$V=V_{\textrm{norm}}$ approximation.

\begin{figure}[h!]
    \includegraphics[width=8.5cm]{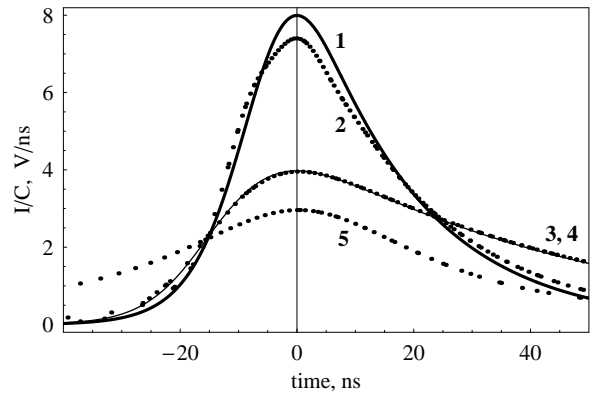}
    \caption{\label{fig:Lg800Va390} Renormalized discharge current for $L_g=800\mu m$ and
    $V_{\textrm{appl}}=390V$:
    (1) $V_{CF}=V_{\textrm{norm}}$ approximation;
    (2) fluid simulation for $2d/\epsilon=1\mu m$;
    (3), (4) fluid simulation and $V_{CF}=V_{CF}(I)$ approximation for $2d/\epsilon=0.1\mu
    m$;
    (5) fluid simulation for $2d/\epsilon=10\mu m$. Time is measured from the moment of
maximum of the currents.}
\end{figure}

\begin{figure}[h!]
    \includegraphics[width=8.5cm]{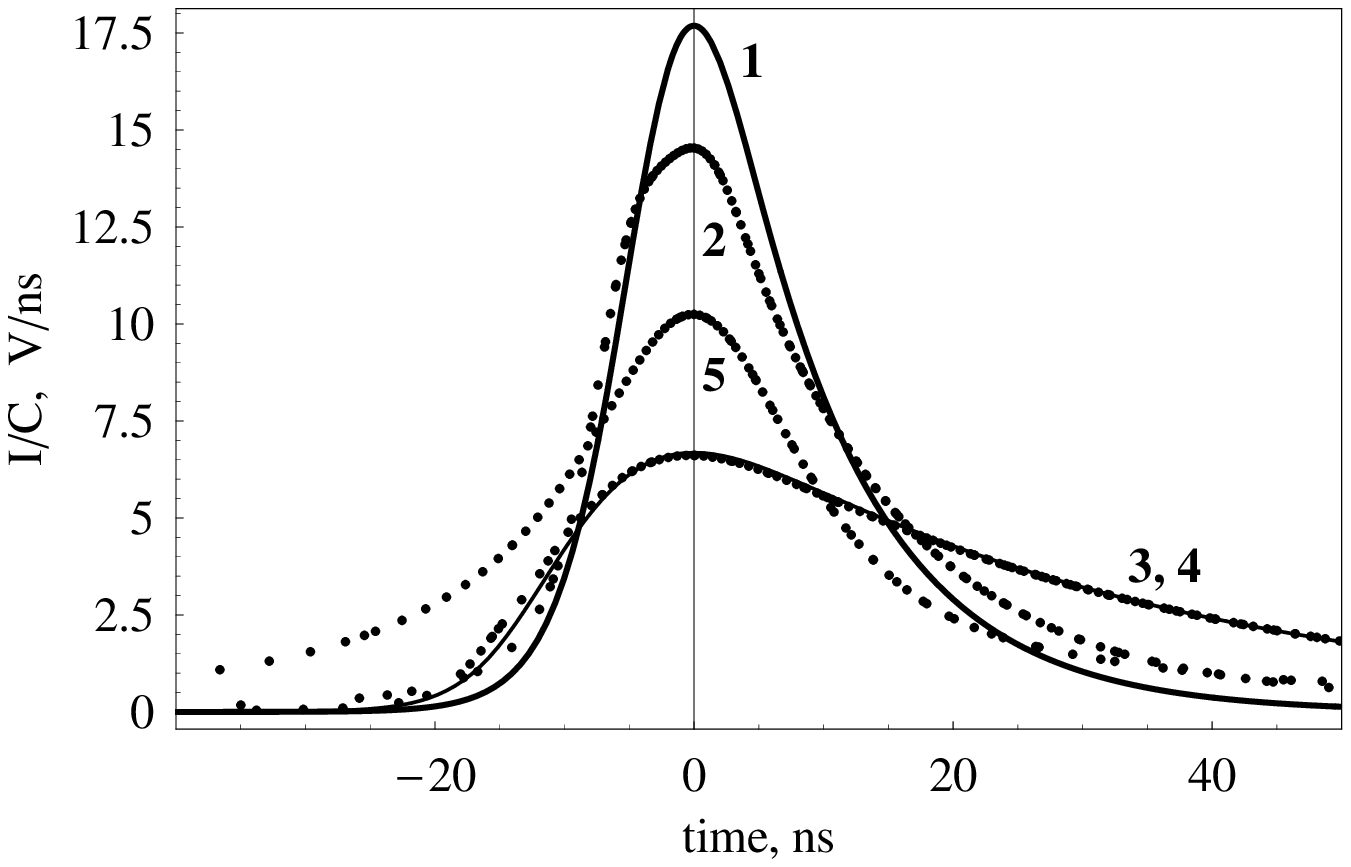}
    \caption{\label{fig:Lg800Va450} The same as in Fig.~\ref{fig:Lg800Va390} but for $L_g=800\mu m$ and
    $V_{\textrm{appl}}=450V$.}
\end{figure}

\begin{figure}[h!]
    \includegraphics[width=8.5cm]{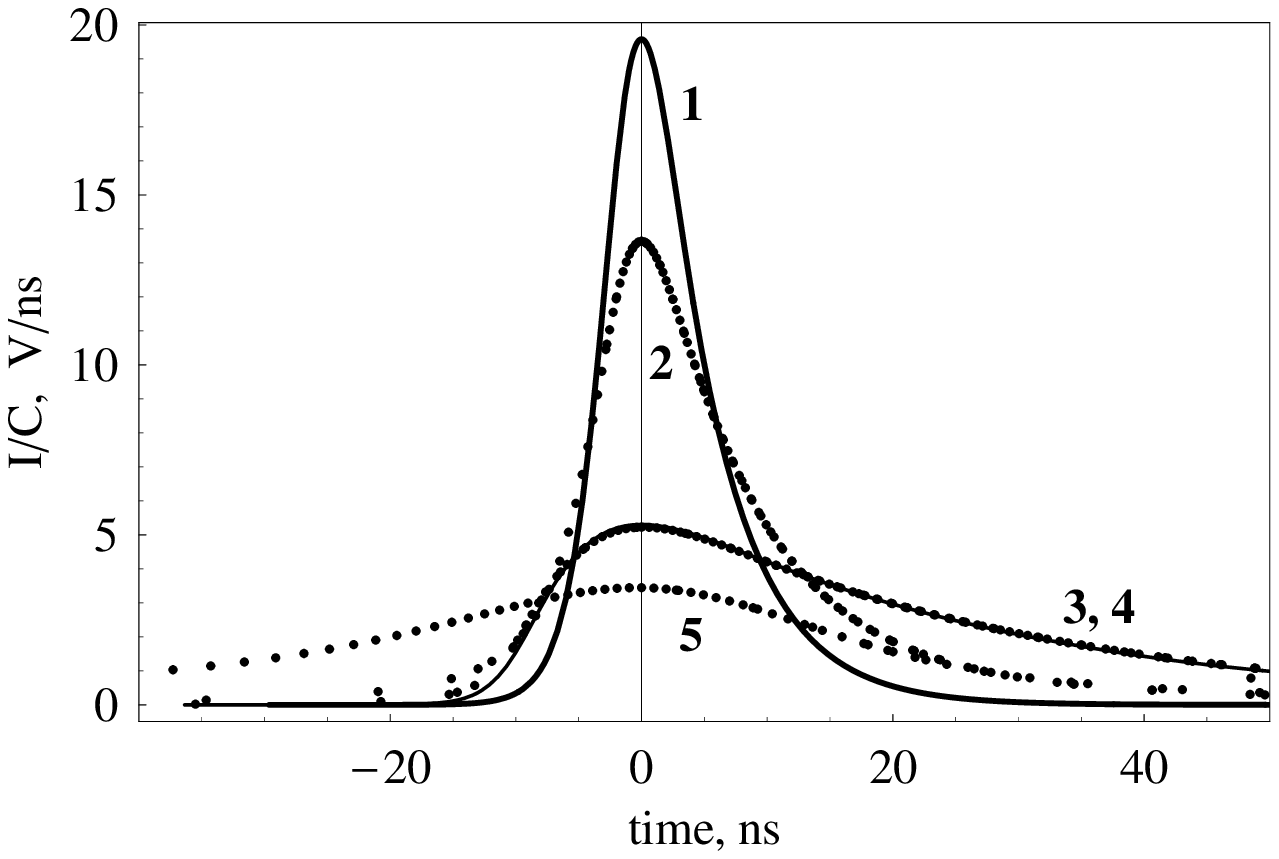}
    \caption{\label{fig:Lg400Va320} The same as in Fig.~\ref{fig:Lg800Va390} but for $L_g=400\mu m$ and
    $V_{\textrm{appl}}=320V$.}
\end{figure}

In Figs.~\ref{fig:Lg800Va390}-\ref{fig:Lg400Va320}
are shown results obtained from our model and from fluid simulations
under different discharge conditions. A striking feature of all
these figures is that the amplitude of the renormalized current
$I/C$ changes by only a factor of $2-3$ when the capacitance $C$
changes by two orders of magnitude (compare curves 2,3 and 5). A
prediction of our model that $I/C$ does not depend on $C$ (see
Eq.~(\ref{eq:jmax})) is qualitatively correct even when the
criterion (\ref{eq:LdielLcf}) of the model applicability is not
satisfied. For curve~5, the effective dielectric thickness
$2d/\epsilon = 10 \mu m$ is greater than the normal CF length
$L_{\textrm{norm}} \approx 6.5 \mu m$, so that the CF is not
quasi-stationary but dynamic!

Understandably, the $V_{CF}=V_{\textrm{norm}}$ approximation
overestimates the amplitude of the discharge current, since
$V_{\textrm{norm}}$ is the minimal voltage on the cathode fall; this
overestimation is more pronounced for shorter gaps.

Another interesting feature of the long-gap discharge dynamics is
that the duration of the current pulse increases with increase of
the gap length (compare Fig.~\ref{fig:Lg400Va320} with
Figs.~\ref{fig:Lg800Va390} and \ref{fig:Lg800Va450}), which is not
at all obvious since the applied voltage is greater in the case of
the longer gap. Qualitatively this discharge property is attributed
to the increase in inertia of the positive column with its length.

Note also that in the case of the large capacitance (abnormal
quasi-stationary CF, see curve 3), the discharge current grows
quickly and falls slowly. In the opposite case of the small
capacitance (dynamic CF, see curve 5), the current grows somewhat
slower than it falls.

Similar to DC discharges, the positive column in barrier discharges
is more efficient source of light (radiation) than the cathode fall.
In the positive column, where the electric field is weak, almost all
the deposited energy goes into excitation of atoms, while in the
cathode fall significant portion of the energy is spent on the ion
heating, and, because the electric field is strong, a considerable
part of the rest goes into ionization of atoms rather than
excitation. It is obvious that the case when the cathode fall is
very abnormal (i.e. when the $C$ is very large) is not optimal for
the light production. The question of the optimal capacitance of
dielectric layers -- i.e. the optimal relationship between
$2d/\epsilon$ and $L_{\textrm{norm}}$ (which determines to what
degree the CF is dynamic) -- should be dealt with in the context of
concrete discharge parameters.

At the end of this section, let us make several remarks regarding
the quasi-neutrality of the plasma in the positive column and
possible generalizations of our model. The quasi-neutrality results
from the fact that Maxwellian relaxation time is much less than the
time of plasma growth in the column (which is of the order of
$\tau_{\textrm{cur}}$). Using Eq.~(\ref{eq:nncfin}), one can
estimate
$$
    \tau_{m}=\frac{\epsilon_0}{e \mu_e n} \sim \frac{\epsilon_0 S}{C L_g \bar{r}(0,E_0)}
    =\frac{2 d}{\epsilon L_g \bar{r}(0,E_0)}.
$$
Since $\bar{r}(0,E_0)=\bar{r}(0,E_*)+\bar{r}(E_*,E_0)>r(E_*) \sim
\tau_{\textrm{cur}}^{-1}$, the quasi-neutrality condition takes the
form
\begin{equation}
            \frac{\tau_{m}}{\tau_{\textrm{cur}}} \sim \frac{2 d}{\epsilon L_g} \ll
            1.
\end{equation}
Our model allows us to include plasma losses due to ambipolar
diffusion to the walls surrounding the discharge volume and due to
recombination by subtracting from the r.h.s. of
Eq.~(\ref{eq:dndtpc}) the terms $\tau^{-1}_{\textrm{dif}} n_{pc}$
($\tau_{\textrm{dif}}$ is the characteristic ambipolar diffusion
time) and $k_{\textrm{r}} n_{pc}^2$ ($k_{\textrm{r}}$ is the
recombination coefficient). Note, however, that the side walls can
strongly influence the breakdown stage of the discharge
\cite{Shvydky2005avalanche} and can even lead to the formation of
near-dielectric-surface discharge rather than the volume one.

\section{\label{sec:conclusion}Conclusion}

The dynamics of a barrier discharge between opposing electrodes
separated by a long gap which is filled with a mixture of noble
gases has been considered. It has been shown that during the
discharge development, there forms a region between the anode and
the cathode fall similar to a DC positive column, where the plasma
density and electric field are uniform but, in contrast to a DC
case, change in time. The mechanism of the positive column
uniformization has an entirely dynamic nature. It is caused by the
ionization processes and has nothing to do with the diffusion.

A simple discharge model allowed us to obtain useful estimates for
the plasma density, number of excitations and ionizations, and power
consumption; it also helped to capture non-trivial trends in the
dependence of the discharge current on discharge parameters.

\bibliography{References}

\begin{thebibliography}{14}
\expandafter\ifx\csname natexlab\endcsname\relax\def\natexlab#1{#1}\fi
\expandafter\ifx\csname bibnamefont\endcsname\relax
  \def\bibnamefont#1{#1}\fi
\expandafter\ifx\csname bibfnamefont\endcsname\relax
  \def\bibfnamefont#1{#1}\fi
\expandafter\ifx\csname citenamefont\endcsname\relax
  \def\citenamefont#1{#1}\fi
\expandafter\ifx\csname url\endcsname\relax
  \def\url#1{\texttt{#1}}\fi
\expandafter\ifx\csname urlprefix\endcsname\relax\def\urlprefix{URL }\fi
\providecommand{\bibinfo}[2]{#2}
\providecommand{\eprint}[2][]{\url{#2}}

\bibitem[{\citenamefont{Meunier et~al.}(1995)\citenamefont{Meunier, Belenguer,
  and Boeuf}}]{Boeuf95}
\bibinfo{author}{\bibfnamefont{J.}~\bibnamefont{Meunier}},
  \bibinfo{author}{\bibfnamefont{P.}~\bibnamefont{Belenguer}},
  \bibnamefont{and} \bibinfo{author}{\bibfnamefont{J.~P.} \bibnamefont{Boeuf}},
  \bibinfo{journal}{{J.\ Appl.\ Phys.}} \textbf{\bibinfo{volume}{78}},
  \bibinfo{pages}{731} (\bibinfo{year}{1995}).

\bibitem[{\citenamefont{Eliasson and Kogelschatz}(1991)}]{Eliasson91}
\bibinfo{author}{\bibfnamefont{B.}~\bibnamefont{Eliasson}} \bibnamefont{and}
  \bibinfo{author}{\bibfnamefont{U.}~\bibnamefont{Kogelschatz}},
  \bibinfo{journal}{{IEEE\ Trans.\ Plasma\ Sci.}}
  \textbf{\bibinfo{volume}{19}}, \bibinfo{pages}{1063} (\bibinfo{year}{1991}).

\bibitem[{\citenamefont{Khudik et~al.}(2003)\citenamefont{Khudik, Nagorny, and
  Shvydky}}]{Khudik2003}
\bibinfo{author}{\bibfnamefont{V.~N.} \bibnamefont{Khudik}},
  \bibinfo{author}{\bibfnamefont{V.~P.} \bibnamefont{Nagorny}},
  \bibnamefont{and} \bibinfo{author}{\bibfnamefont{A.}~\bibnamefont{Shvydky}},
  \bibinfo{journal}{{J.\ Appl.\ Phys.}} \textbf{\bibinfo{volume}{94}},
  \bibinfo{pages}{6291} (\bibinfo{year}{2003}).

\bibitem[{\citenamefont{{Yu. P. Raizer}}(c1991)}]{RaizerBook}
\bibinfo{author}{\bibnamefont{{Yu. P. Raizer}}}, \emph{\bibinfo{title}{Gas
  Discharge Physics}} (\bibinfo{publisher}{Berlin ; New York :
  Springer-Verlag}, \bibinfo{year}{c1991}).

\bibitem[{\citenamefont{Boeuf}(2003)}]{Boeuf2003review}
\bibinfo{author}{\bibfnamefont{J.~P.} \bibnamefont{Boeuf}},
  \bibinfo{journal}{{J.\ Phys.\ D:\ Appl.\ Phys.}}
  \textbf{\bibinfo{volume}{36}}, \bibinfo{pages}{R53} (\bibinfo{year}{2003}).

\bibitem[{\citenamefont{Weber}(2001)}]{Weber2001patent}
\bibinfo{author}{\bibfnamefont{L.}~\bibnamefont{Weber}},
  \emph{\bibinfo{title}{{\it US Patent} \rm 6184848b1}} (\bibinfo{year}{2001}).

\bibitem[{\citenamefont{Schermerhorn et~al.}(2000)\citenamefont{Schermerhorn,
  Anderson, Levison, Hammon, Kim, Park, Ryu, Shvydky, and
  Sebastian}}]{JerrySID2000}
\bibinfo{author}{\bibfnamefont{J.~D.} \bibnamefont{Schermerhorn}},
  \bibinfo{author}{\bibfnamefont{E.}~\bibnamefont{Anderson}},
  \bibinfo{author}{\bibfnamefont{D.}~\bibnamefont{Levison}},
  \bibinfo{author}{\bibfnamefont{C.}~\bibnamefont{Hammon}},
  \bibinfo{author}{\bibfnamefont{J.~S.} \bibnamefont{Kim}},
  \bibinfo{author}{\bibfnamefont{B.~Y.} \bibnamefont{Park}},
  \bibinfo{author}{\bibfnamefont{J.~H.} \bibnamefont{Ryu}},
  \bibinfo{author}{\bibfnamefont{A.}~\bibnamefont{Shvydky}}, \bibnamefont{and}
  \bibinfo{author}{\bibfnamefont{A.}~\bibnamefont{Sebastian}},
  \bibinfo{journal}{{SID 00 Digest}} \textbf{\bibinfo{volume}{31}},
  \bibinfo{pages}{106} (\bibinfo{year}{2000}).

\bibitem[{\citenamefont{Kawai et~al.}(2004)\citenamefont{Kawai, Tachibana, Oh,
  Asai, Kikuchi, and Sakamoto}}]{Kawai2004}
\bibinfo{author}{\bibfnamefont{S.}~\bibnamefont{Kawai}},
  \bibinfo{author}{\bibfnamefont{K.}~\bibnamefont{Tachibana}},
  \bibinfo{author}{\bibfnamefont{J.}~\bibnamefont{Oh}},
  \bibinfo{author}{\bibfnamefont{H.}~\bibnamefont{Asai}},
  \bibinfo{author}{\bibfnamefont{N.}~\bibnamefont{Kikuchi}}, \bibnamefont{and}
  \bibinfo{author}{\bibfnamefont{S.}~\bibnamefont{Sakamoto}},
  \bibinfo{journal}{Int. Display Workshop \textbf{IDW'04}} pp.
  \bibinfo{pages}{1059--1062} (\bibinfo{year}{2004}).

\bibitem[{\citenamefont{Punset et~al.}(1999)\citenamefont{Punset, Cany, and
  Boeuf}}]{Boeuf99}
\bibinfo{author}{\bibfnamefont{C.}~\bibnamefont{Punset}},
  \bibinfo{author}{\bibfnamefont{S.}~\bibnamefont{Cany}}, \bibnamefont{and}
  \bibinfo{author}{\bibfnamefont{J.~P.} \bibnamefont{Boeuf}},
  \bibinfo{journal}{{J.\ Appl.\ Phys.}} \textbf{\bibinfo{volume}{86}},
  \bibinfo{pages}{124} (\bibinfo{year}{1999}).

\bibitem[{\citenamefont{Rauf and Kushner}(1999)}]{Kushner99I}
\bibinfo{author}{\bibfnamefont{S.}~\bibnamefont{Rauf}} \bibnamefont{and}
  \bibinfo{author}{\bibfnamefont{M.~J.} \bibnamefont{Kushner}},
  \bibinfo{journal}{{J.\ Appl.\ Phys.}} \textbf{\bibinfo{volume}{85}},
  \bibinfo{pages}{3460} (\bibinfo{year}{1999}).

\bibitem[{\citenamefont{Khudik et~al.}(2005)\citenamefont{Khudik, Nagorny, and
  Shvydky}}]{Khudik2005SID}
\bibinfo{author}{\bibfnamefont{V.~N.} \bibnamefont{Khudik}},
  \bibinfo{author}{\bibfnamefont{V.~P.} \bibnamefont{Nagorny}},
  \bibnamefont{and} \bibinfo{author}{\bibfnamefont{A.}~\bibnamefont{Shvydky}},
  \bibinfo{journal}{{J.\ SID.}} \textbf{\bibinfo{volume}{13}},
  \bibinfo{pages}{147} (\bibinfo{year}{2005}).

\bibitem[{\citenamefont{Khudik et~al.}()\citenamefont{Khudik, Shvydky, Nagorny,
  and Theodosiou}}]{Khudik2005Images}
\bibinfo{author}{\bibfnamefont{V.~N.} \bibnamefont{Khudik}},
  \bibinfo{author}{\bibfnamefont{A.}~\bibnamefont{Shvydky}},
  \bibinfo{author}{\bibfnamefont{V.~P.} \bibnamefont{Nagorny}},
  \bibnamefont{and} \bibinfo{author}{\bibfnamefont{C.~E.}
  \bibnamefont{Theodosiou}}, \bibinfo{note}{to be published in 4th Triennial
  Special Issue of the IEEE Transactions on Plasma Science "Images in Plasma
  Science", April 2005.}

\bibitem[{\citenamefont{Shvydky
  et~al.}(2004{\natexlab{a}})\citenamefont{Shvydky, Khudik, Nagorny, and
  Theodosiou}}]{shvydky2004}
\bibinfo{author}{\bibfnamefont{A.}~\bibnamefont{Shvydky}},
  \bibinfo{author}{\bibfnamefont{V.~N.} \bibnamefont{Khudik}},
  \bibinfo{author}{\bibfnamefont{V.~P.} \bibnamefont{Nagorny}},
  \bibnamefont{and} \bibinfo{author}{\bibfnamefont{C.~E.}
  \bibnamefont{Theodosiou}}, \emph{\bibinfo{title}{Dynamics of the breakdown in
  a discharge gap at high overvoltages}}, \bibinfo{howpublished}{GEC'57, Sept.
  26-29} (\bibinfo{year}{2004}{\natexlab{a}}).

\bibitem[{\citenamefont{Shvydky
  et~al.}(2004{\natexlab{b}})\citenamefont{Shvydky, Khudik, and
  Nagorny}}]{Shvydky2005avalanche}
\bibinfo{author}{\bibfnamefont{A.}~\bibnamefont{Shvydky}},
  \bibinfo{author}{\bibfnamefont{V.~N.} \bibnamefont{Khudik}},
  \bibnamefont{and} \bibinfo{author}{\bibfnamefont{V.~P.}
  \bibnamefont{Nagorny}}, \bibinfo{journal}{{J.\ Phys.\ D:\ Appl.\ Phys.}}
  \textbf{\bibinfo{volume}{37}}, \bibinfo{pages}{2996}
  (\bibinfo{year}{2004}{\natexlab{b}}).

\end{thebibliography}


\end{document}